\documentclass[aps,prb,twocolumn,superscriptaddress,floatfix]{revtex4}

\usepackage{graphicx}
\begin{document}  

\title{Comparison of three different self-interaction corrections for an 
exactly solvable model system}
\author{Daniel Vieira}
\affiliation{Departamento de F\'{\i}sica e Inform\'atica,
Instituto de F\'{\i}sica de S\~ao Carlos\\ 
Universidade de S\~ao Paulo,
Caixa Postal 369, S\~ao Carlos, 13560-970 SP, Brazil}
\author{K. Capelle}
\email{capelle@ifsc.usp.br}
\affiliation{Departamento de F\'{\i}sica e Inform\'atica,
Instituto de F\'{\i}sica de S\~ao Carlos\\
Universidade de S\~ao Paulo,
Caixa Postal 369, S\~ao Carlos, 13560-970 SP, Brazil}
\affiliation{Theoretische Physik, Freie Universit\"at Berlin, D-14195 Berlin,
Germany}

\date{\today}

\begin{abstract}
A systematic comparison of three approximate self-interaction corrections 
(SICs), Perdew-Zunger SIC, Lundin-Eriksson SIC and extended Fermi-Amaldi SIC, 
is performed for a model Hamiltonian whose exact many-body solution and exact 
local-density approximation (LDA) are known. For each of the three proposals 
we compare its implementation only for the potential, only for the energy 
({\em i.e.}, a post-LDA evaluation of the SIC energy), to none of them ({\em 
i.e.}, a standard LDA calculation) and to both. Each 
of the resulting 10 permutations of methodologies is applied to 420 Hubbard 
chains differing in size, particle number and interaction strength. A
statistical analysis of the resulting data set reveals trends and permits to
assess the performance of each methodology. Overall, but not in each 
individual case, a post-LDA application of Perdew-Zunger SIC emerges as the
recommended methodology.
\end{abstract}

\maketitle

\newcommand{\be}{\begin{equation}}
\newcommand{\ee}{\end{equation}}
\newcommand{\bea}{\begin{eqnarray}}
\newcommand{\eea}{\end{eqnarray}}
\newcommand{\bi}{\bibitem}

\renewcommand{\r}{({\bf r})}
\newcommand{\rp}{({\bf r'})}

\newcommand{\ua}{\uparrow}
\newcommand{\da}{\downarrow}
\newcommand{\la}{\langle}
\newcommand{\ra}{\rangle}
\newcommand{\dg}{\dagger}

\section{Introduction}
\label{intro}

Density-functional theory (DFT)\cite{kohnrmp,dftbook,parryang} is advancing 
at a rapid pace, driven by the demand of ever more accurate 
electronic-structure calculations. To meet this demand, ever more 
sophisticated density functionals are being constructed. 
A simultaneous, and often contradictory, demand is the applicability to 
larger and more complex systems, met by computational and methodological
advances in the implementation of density functionals.
In this context, the comparison of different approximate density functionals
and different modes of their implementation is of key importance.
The present paper provides such a comparison for the case of
self-interaction corrected functionals. 

In Sec.~\ref{siesic} we recall three different self-interaction corrections.
In Sec.~\ref{methods} we discuss four different implementations, selected
specifically to be able to compare the effect of self-interaction corrections
on the potential and on the energy, in an unbiased way. 
In Sec.~\ref{comparisons} we report results from a statistical analysis 
of hundreds of calculations with each of the possible permutations of 
correction and implementation, for an exactly solvable model system whose
exact many-body ground-state energy is known. 
Sec.~\ref{conclusions} contains our conclusions.

\section{Self-interaction error and self-interaction
corrections}
\label{siesic}

One electron does not interact with itself. Simple as it is, this fact is 
at the heart of much trouble in many-body and electronic-structure theory. 
Traditional density functionals, such as the local density, 
local spin-density and generalized-gradient approximations (GGAs), as well 
as many more recently developed functionals, such as hybrids and meta GGAs,
do have a spurious self-interaction error (SIE), and as a consequence predict 
a nonzero interaction energy even for a single electron. 
The {\em exact} exchange energy of one electron exactly cancels its Hartree 
energy, so that the Hartree + exchange energy of methods employing exact 
exchange is self-interaction free, but the correlation energy remains subject 
to self interaction even in methods employing exact exchange. Consequences of 
the SIE of approximate functionals are multifarious, 
and include wrong asymptotics of approximate exchange-correlation ($xc$)
potentials, energy gaps, ionization energies, electron affinities and
transition-metal magnetic moments.

Already in 1934, Fermi and Amaldi (FA)\cite{fapaper} realized that 
Thomas-Fermi calculations, which use a local-density approximation for the 
noninteracting kinetic energy and approximate the full interaction energy by 
the Hartree energy,
\be
E_H[n] = \frac{1}{2} \int d^3 r \int d^3 r'\, \frac{n\r n\rp}{|{\bf r - r'}|},
\ee
suffer from a large SIE, and proposed a first 
self-interaction correction (SIC), of the form
\be
E_H^{FASIC}[n] = \left(1-\frac{1}{N}\right) E_H[n]
=E_H[n] -\frac{1}{N} E_H[n].
\label{fasic}
\ee
Equation (\ref{fasic}) is clearly a very crude approximation, violating, 
among other things, the requirement of size-consistency, but it does eliminate 
the SIE of $E_H[n]$ for a one-electron system ($N=1$). 

After the development of modern density-functional 
theory, the question of how to eliminate the SIE posed itself anew. One 
way to carry over to modern Kohn-Sham DFT the simplicity of the FA 
approach is what we call extended-Fermi-Amaldi SIC (EFASIC), where the FA 
factor is applied not just to the Hartree energy, but to the sum of 
Hartree, exchange and correlation energies, 
\bea
E_{xc}^{EFASIC}[n_\ua, n_\da]= E_{xc}^{approx}[n_\ua, n_\da]
\nonumber \\
-\frac{1}{N} (E_H[n] + E_{xc}^{approx}[n_\ua, n_\da]),
\label{efasic}
\eea
where $E_{xc}^{approx}[n_\ua, n_\da]$ is an approximation, {\em e.g.} LDA or 
GGA, to the exact exchange-correlation functional.

A very successful proposal for a more sophisticated SIC, made 
in 1981 by Perdew and Zunger (PZ),\cite{pzpaper} is
\bea
E_{xc}^{PZSIC}[n_\ua,n_\da]=E_{xc}^{approx}[n_\ua,n_\da] 
 \nonumber \\
- \sum_k^{occ} \sum_{\sigma=\uparrow,\downarrow} 
\left(E_H[n_{k\sigma}] + E^{approx}_{xc}[n_{k\sigma},0]\right),
\label{pzsic}
\eea
where $n_\sigma\r=\sum^{occ}_{k} n_{k\sigma}\r$, 
$n_{k\sigma}\r = |\varphi_{k\sigma}\r|^2$, and 
$\varphi_{k\sigma}\r$ are the Kohn-Sham (KS) orbitals.

In PZSIC, the SIE is subtracted on a orbital by orbital basis. Evidently,
for a one-electron density $E_{xc}^{PZSIC}[n^{(1)}]=-E_H[n^{(1)}]$.
The PZSIC functional depends on the orbitals $\varphi_{k\sigma}$, so that 
its minimization with respect to the density must employ the 
optimized-effective potential (OEP) method
or one of its simplifications.\cite{kli,kummelperdew,yangwu}
In practice, however, the minimization is usually performed with respect 
to the orbital densities, and 
\bea
v_{s,k\sigma}^{PZSIC}({\bf r})=v_{ext}\r + v_H[n]\r +v_{xc,\sigma}^{approx}[n_\sigma,n_{\bar{\sigma}}]({\bf r})
\nonumber \\
- v_H[n_{k\sigma}]\r -v_{xc,\sigma}^{approx}[n_{k\sigma},0]\r
\\
=:v_{ext}\r + v_H[n]\r + v_{xc,k\sigma}^{PZSIC}[n_\sigma,n_{\bar{\sigma}}]({\bf r})
\eea
is adopted as corresponding single-body potential.
This potential is constructed from the density of {\em all} electrons.

The PZSIC approach, simplified by minimizing with respect to orbital densities
instead of total densities, has been very succesful in removing the 
one-electron SIE in 
solids.\cite{temmerman} Results from proper minimization with respect to the 
densities, by means of the OEP, are reported in, {\em e.g.},
Refs.~\onlinecite{pzOEP1,pzOEP2} for atomic systems. PZSIC has become 
so popular that frequently the abbreviation SIC is used as synonymous with 
PZSIC, but it has been repeatedly noted that the PZ proposal is not the only 
possibility for removing the SIE.\cite{cortona,unger,voskowilk,lepaper}

An innovative proposal for an alternative self-interaction correction was 
put forward in 2001 by Lundin and Eriksson (LE).\cite{lepaper}
The LE proposal attempts to construct an effective potential that, when acting 
on the orbital of one electron, is constructed only from the density of the 
{\em other} electrons. In practice, this is done by  introducing an 
orbital-dependent effective potential, $v_{s,k \sigma}\r$, determined by 
subtracting from the full density $n\r$ the partial density of the orbital 
the potential acts on,\cite{lepaper,novak}
\bea
v_{s,k \sigma}^{LESIC}({\bf r})=
v_{ext}\r+ v_H[n-n_{k\sigma}]({\bf r})
\nonumber \\
 + v^{approx}_{xc}[n_\sigma-n_{k\sigma},n_{\bar{\sigma}}]({\bf r})
\label{lesicV}
\\
=v_{ext}\r+ v_H[n]\r - v_H[n_{k\sigma}]({\bf r})
\nonumber \\
 +  v^{approx}_{xc}[n_\sigma-n_{k\sigma},n_{\bar{\sigma}}]({\bf r})
\\
=:v_{ext}\r+ v_H[n]\r +  v_{xc,k\sigma}^{LESIC}[n_\sigma,n_{\bar{\sigma}}]({\bf r}).
\label{lesicV2}
\eea
In this way, the approximate effective potential acting on a given orbital,
$v_{s, k\sigma}^{LESIC}({\bf r})$, is constructed only from the density 
arising from the other orbitals. Clearly, for a one-particle system this 
approach also correctly zeroes the interaction contribution to the effective 
potential. 

The LE proposal for a corrected effective potential is accompanied by a 
similar expression for the exchange-correlation energy,\cite{lepaper,seo}
\bea
E_{xc}^{LESIC}[n_\ua,n_\da]=-\sum_k^{occ} \sum_{\sigma=\uparrow,\downarrow} 
E_H[n_{k \sigma}]
\nonumber \\
+ \sum_k^{occ} \sum_{\sigma=\uparrow,\downarrow}
\int dr^3 \, n_{k \sigma}\r
e_{xc}^{approx}[n_\sigma-n_{k \sigma}, n_{\bar{\sigma}}].
\label{lesicE}
\eea
We note that the change from the LESIC to the PZSIC approach can 
be affected by substituting
\be 
e_{xc}^{approx}[n_\sigma -n_{k \sigma}, n_{\bar{\sigma}}]
\to
e_{xc}^{approx}[n_\sigma, n_{\bar{\sigma}}]
-e_{xc}^{approx}[n_{k \sigma},0]
\ee
and
\be
v_{xc}^{approx}[n_\sigma-n_{k\sigma},n_{\bar{\sigma}}]
\to
v_{xc,\sigma}^{approx}[n_\sigma,n_{\bar{\sigma}}]
- v_{xc,\sigma}^{approx}[n_{k\sigma},0]
\ee
in the $xc$ energy density and potential of former. For the Hartree potential,
the substitution $v_H[n-n_{k\sigma}] \to v_H[n] - v_H[n_{k\sigma}]$
is an identity.

The LE energy expression (\ref{lesicE}) is constructed in analogy to 
Eq.~(\ref{lesicV2}), but its functional derivative is not the LE $xc$ potential 
in Eq.~(\ref{lesicV2}). Rather, both are separate constructions. This gives rise
to a certain ambiguity in how to implement the LE approach. 
Sec.~\ref{methods} we therefore discuss various alternative implementations 
of Eqs.~(\ref{lesicV2}) and (\ref{lesicE}).

In the original LE work\cite{lepaper} it was argued that the LE approach 
should be superior to the PZ approach because it removes the SIE of the 
potential, which, according to Ref.~\onlinecite{lepaper}, remains in the 
PZ approach. It has been
objected\cite{perdewoverview} that the LE proposal cannot be right because 
it would even correct the hypothetical exact functional, for which the PZ 
approach correctly reduces to zero.

While we agree with this objection as a matter of principle, we feel that 
in practice the key issue is not only what the correction does to the 
hypothetical exact functional, but also what it does to actually available 
approximate functionals. If such functionals were consistently improved by 
a correction, few workers would refrain from using this correction in 
practice, only because it overcorrects the hypothetical exact functional.

Interestingly, there are hints in the literature that approximate $xc$
functionals do indeed benefit more from LESIC than from PZSIC. 
Novak et al.\cite{novak} compare the performance of LSDA, LSDA+PZSIC and 
LSDA+LESIC in the calculation of hyperfine parameters, and find that LESIC
significantly improves agreement with experiment, relative to PZSIC. 
(See also Ref. \onlinecite{joice} for a successful application of LESIC to 
hyperfine parameters.) In a separate study, Friis et al.\cite{friis} compared 
density distributions predicted for the Mg crystal with experimental 
electron diffraction data, and observed that LESIC produces better core 
and valence densities than PZSIC.

The available data are too limited, however, to already conclude that 
LESIC is definitely superior, in practice, to PZSIC, in particular since,
as a matter of principle, LESIC cannot be correct. It therefore becomes
an important task to systematically investigate the performance of both 
approaches. The present paper is a first step towards this task, providing
a systematic analysis of the performance of PZSIC, LESIC and EFASIC for 
a well controlled model system whose exact solution is known. 

\section{Implementation}
\label{methods}

Formally, the $xc$ potential corresponding to a given approximation to 
$E_{xc}[n_\sigma,n_{\bar{\sigma}}]$ is
\be
v_{xc,\sigma}^{approx}[n_\sigma,n_{\bar{\sigma}}]({\bf r})=
\frac{\delta E_{xc}^{approx}[n_\sigma,n_{\bar{\sigma}}]}{\delta n_\sigma({\bf r})}.
\ee
We have just seen, however, that by construction the LE $xc$ potential is 
not the derivative of the LE $xc$ energy, but a separate construction. 
Strictly speaking, the same is true for most common implementations of
PZSIC, where the implemented potential is the orbital derivative, and not
the density derivative of the energy. This ambiguity suggests that a complete
test of each proposal involves four different permutations of methodologies.

Our notation for these is APPV-APPE, where APPV represents the approximation
made for the potential entering the KS equations, and APPE the 
approximation made for the energy functional. The simplest possibility is thus
LDA-LDA, which is a standard LDA calculation using the L(S)DA
potential and energy.\cite{footnote1} The second scheme, denoted
LDA-SIC, consists of a self-consistent LDA calculation, followed by
a single evaluation of the SIE corrected energy functional on the LDA
densities and orbitals. This strategy is also known as post-LDA implementation
of SIC. By construction, it tests the performance of the SIC for the $xc$
energy. The inverse possibility, which we label SIC-LDA, is a self-consistent
SIC calculation, followed by a single evaluation of the uncorrected LDA 
energy functional on the corrected densities. This procedure tests the
performance of the employed $xc$ correction for the potentials. Finally, 
SIC-SIC refers to a calculation using the SIC under study both in the
potential and the energy. 
Starting from LDA-LDA, each of the other three possibilities can be implemented
for each of the three SICs described in the previous section, resulting in
a total of ten combinations of methodologies.

In the self-consistent implementation of the PZSIC potential obtained from
the orbital derivative of the PZ energy and of the LESIC potential, which are
required in schemes SIC-LDA and SIC-SIC, we face the additional difficulty 
that these potentials are not common multiplicative potentials for all KS
orbitals, but are orbital-specific, {\em i.e.} different for each orbital they 
act on. Such orbital-specific potentials, $v_k[\{\varphi_j[n]\}]\r$, are not 
proper Kohn-Sham potentials, although they are still density functionals.

A common multiplicative potential for all orbitals can be generated from 
the PZ and LE energy functionals by means of the OEP, but in the case of LESIC
the potential resulting from the energy functional (\ref{lesicE}) would have 
no relation to the LE potential (\ref{lesicV2}), which, as we have seen above, 
is a separate construction. Hence, implementation via the OEP would implicitly
test the proposed energy correction (\ref{lesicE}), not the proposed potential 
correction (\ref{lesicV2}), required in the schemes SIC-LDA and SIC-SIC.

A way to directly compare the orbital-specific PZ and LE potentials is to 
average the $xc$ potentials over all orbitals, weighted with the contribution 
each orbital makes to the full spin density, according to\cite{slater}
\bea
\bar{v}_{xc,\sigma}^{PZSIC}[n_\sigma,n_{\bar{\sigma}}]({\bf r})=
v_{xc,\sigma}^{approx}[n_\sigma,n_{\bar{\sigma}}]({\bf r})
\nonumber \\
-\sum_k^{occ}(v_H[n_{k\sigma}]\r
+v_{xc}^{approx}[n_{k\sigma},0]\r)
\frac{ n_{k \sigma}({\bf r})}{n_\sigma({\bf r})}
\label{averagepz} 
\eea
and
\bea
\bar{v}_{xc, \sigma}^{LESIC}[n_\sigma,n_{\bar{\sigma}}]({\bf r}) = 
\nonumber \\
- \sum_k^{occ} (v_H[n_{k\sigma}]\r  
- v_{xc}^{approx}[n_\sigma-n_{k\sigma},n_{\bar{\sigma}}]\r)  
\frac{ n_{k \sigma}({\bf r})}{n_\sigma({\bf r})}.
\label{averagele}
\eea
The result are average multiplicative potentials arising from the
orbital-specific LE or PZ potential. Of course, some information is lost
in the averaging, but since our aim is to compare LE and PZ among
each other, the errors of the averaging procedure are of secondary importance,
as long as all functionals are treated in the same way. 

We note that in the case of the PZ correction the same averaged potential is
also obtained by applying the OEP algorithm to the PZ $xc$ functional, 
which in the first step leads to
\bea
v_{xc,\sigma}^{PZSIC}[n_\sigma,n_{\bar{\sigma}}]({\bf r}) =
\frac{\delta E_{xc}^{PZSIC}[n_\sigma,n_{\bar{\sigma}}]}{\delta n_\sigma({\bf r})} 
\\
= v_{xc,\sigma}^{approx}[n_\sigma,n_{\bar{\sigma}}]({\bf r}) 
- \sum_{k\tau}^{occ}\int d^3 r' \left(v_H[n_{k \tau}]\r
\right. \nonumber \\ \left. 
+v_{xc}^{approx}[n_{k \tau},0]\r \right)
\frac{\delta n_{k \tau}({\bf r'})}{\delta n_\sigma({\bf r})}
\eea 
and then making the Slater approximation\cite{slater}
\begin{equation}
\label{eq12} \frac{\delta n_{k \tau}({\bf r'})}{\delta
n_\sigma({\bf r})} \approx \frac{ n_{k \tau}({\bf r'})}{
n_\sigma({\bf r})} \delta({\bf r} - {\bf r'})\delta_{\sigma \tau}
\end{equation}
for the functional derivative. In this sense, our SIC-LDA and SIC-SIC
implementations of PZSIC are approximate OEP calculations. As explained above, 
the same is not true for LESIC, where the OEP would produce the local 
$xc$ potential coming from the LE $xc$ energy (\ref{lesicE}), which is 
not the same as the LE $xc$ potential in Eq.~(\ref{lesicV2}) or its 
average (\ref{averagele}). In the present 
context, it is the averaging prescription which must be employed to test 
the LE proposal for the potential, not the OEP.

Finally, the question poses itself by what standard we are to measure the 
performance of a SIC. Comparison to experiment is hard because the SIE itself 
is not an experimental observable, and analysis of fine differences between 
computational approaches relative to experiment is complicated by effects such 
as relativity, finite temperature, zero-point vibrations, sample purity, etc. 
that are of no direct relevance to the calculation, but automatically included 
in the experimental value. Even on the purely computational side, the issue is 
thorny, because the approximate SIC is applied to a functional, such as LDA, 
that is itself approximate, and it is not always easy to disentangle the error 
of the functional from that of its correction, or from that of approximations
such as the sphericalization of atomic densities often used in PZSIC 
calculations. Moreover, calculations for real systems involve basis sets or 
numerical meshes, whose adequateness must be carefully tested, and which may 
introduce further residual uncertainties in the calculations. 

For all these reasons, we perform our analysis using as theoretical laboratory 
a very well controlled model Hamiltonian: the one-dimensional Hubbard model
\be
\hat{H} = -t \sum_{i, \sigma}^L \left(c_{i
\sigma}^{\dagger}c_{i+1, \sigma} + \textrm{H.c.} \right) +
U\sum_{i}^L c_{i \uparrow}^{\dagger}c_{i \uparrow} c_{i
\downarrow}^{\dagger}c_{i \downarrow}.
\ee
This model, which may be considered a special case of the PPP model popular
in quantum chemistry,\cite{ppp} describes electrons on a one-dimensional
lattice of size $L$, interacting with on-site interaction $U$, and hopping
from one site $i$ to the next with amplitude $t$. Occupation of each site is 
limited to two particles, necessarily of opposite spin. Like-spin particles
do not interact in the most common form of the model.
The basic Hohenberg-Kohn and Kohn-Sham theorems of DFT hold for this model, 
too, once the density $n\r$ is replaced by the on-site occupation
number\cite{gs,gsn} 
\be
n\r = \sum_\sigma n_\sigma\r \rightarrow n_i = \sum_{\sigma} n_{i\sigma}
= \sum_{\sigma} \la c_{i \sigma}^{\dagger}c_{i \sigma}\ra.
\ee
In terms of this variable, local-density and spin-density approximations for 
Hubbard chains and rings have been constructed,\cite{balda,mott,oxford,coldatoms} and can be employed in the usual way.

What makes this model attractive as a theoretical laboratory for DFT is that
its exact many-body solution is known in the thermodynamic limit ($L\to 
\infty$, $n_i=n=const$),\cite{lieb} and an exact numerical solution is possible 
for small systems ($L\sim 14$ on a workstation, for arbitrary density 
distributions $n_i$). As a consequence of the availability of the exact 
solution for the infinite homogeneous system, the exact LDA is 
known, too.\cite{footnote2} Moreover, a complete basis consists of two 
spin-orbitals per site, so that we can always work at the basis-set limit.
Another welcome feature, which we will exploit 
below, is that the model allows to continuously vary the 
interaction strength both in the exact and in the DFT calculations, which 
simulates the behavior of weakly and strongly interacting systems in {\em 
ab initio} calculations. 

\section{Results}
\label{comparisons}

In this section we compare the three self-interaction corrections EFASIC, 
LESIC and PZSIC, in the LDA-LDA, LDA-SIC, SIC-LDA and SIC-SIC schemes,
using the orbital (Slater-type) average of Eqs.~(\ref{averagepz}) and 
(\ref{averagele}) to deal with
the orbital-specific PZ and LE potentials in the schemes SIC-LDA and SIC-SIC.
All calculations are done for finite chains of different lengths $L$, 
particle numbers $N$ and interaction strengths $U$. Variation of these
parameters for each combination of methodology produces thousands of
data, which we analyze statistically by reporting the percentage 
root-mean-square (rms) error,

\begin{equation}
rms=100 \sqrt{\frac{1}{N_S}\sum_{j=1}^{N_S}
\left(\frac{E^j_{approx}-E^j_{exact}}{E^j_{exact}}\right)^2},
\label{rms}
\end{equation}
of each approximation relative to numerically exact many-body energies for 
the same set of model parameters. $N_S$ is the number of systems in the data 
set, labelled by $j$.

\begin{figure}
\includegraphics[height=65mm,width=80mm,angle=-0]{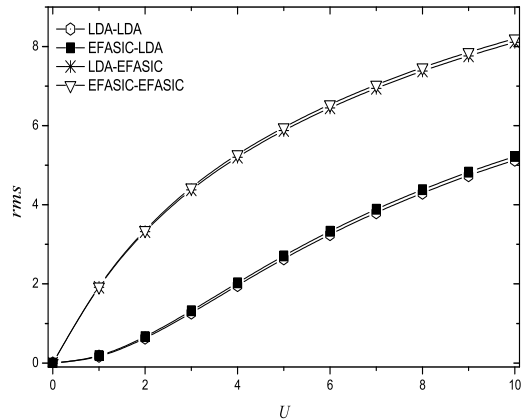}
\caption{\label{sicfig1} Root-mean-square error, in percent, of the 
different implementations of the Extended Fermi-Amaldi self-interaction
correction (EFASIC), relative to the exact many-body ground-state energy.
$N_s=42$ systems are considered at each $U/t$.}
\end{figure}

\begin{figure}
\includegraphics[height=65mm,width=80mm,angle=-0]{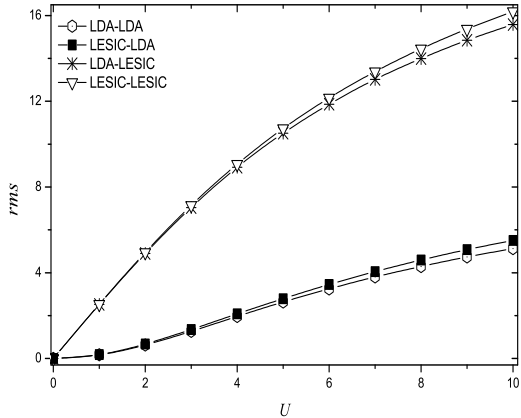}
\caption{\label{sicfig2} Root-mean-square error, in percent, of the
different implementations of the Lundin-Eriksson self-interaction
correction (LESIC), relative to the exact many-body ground-state energy.
$N_s=42$ systems are considered at each $U/t$.}
\end{figure}

\begin{figure}
\includegraphics[height=65mm,width=80mm,angle=-0]{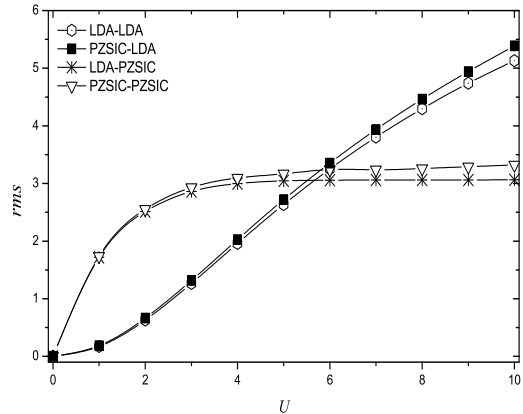}
\caption{\label{sicfig3} Root-mean-square error, in percent, of the
different implementations of the Perdew-Zunger self-interaction
correction (PZSIC), relative to the exact many-body ground-state energy.
$N_s=42$ systems are considered at each $U/t$.}
\end{figure}

Our results are summarized in Figs.~\ref{sicfig1} to \ref{sicfig3} and in
Table~\ref{sictable}. Each of the three figures compares the rms error
defined in Eq.~(\ref{rms}) as a function of interaction strength $U$, ranging 
from the noninteracting system $U=0$, over weakly $U\sim 2t$, strongly 
$U\sim6t$ to very strongly, $U\sim10 t$, interacting systems. For each value 
of $U/t$, our sample consists of $N_s=42$ systems of different sizes, particle
number, and densities ($L=3,4,5..14$ and $N=2,4,6,..12$, with $N/L<1$).

We can analyze these figures from two different points of view. First, 
comparing the three corrections, we find that PZSIC consistently performs much
better than LESIC or EFASIC. The scale on the vertical axis of the three 
figures is not the same, to permit a better visualization of the differences 
between
the various implementation schemes, but the values of the rms error 
are directly comparable, and leave little doubt about the superiority of 
PZSIC for these systems. Somewhat surprisingly, the very simple EFASIC 
still performs a little better than the more sophisticated LESIC. Of course,
our conclusions are based only on ground-state energies, but since the 
difference in performance for the energy is quite large, and since many
other observables are obtained from energies, we expect that PZSIC should
be superior to the others in most applications.

Second, comparing modes of implementation, we find that they fall in two 
groups, distinguished by the energy functional: The schemes LDA-LDA and 
SIC-LDA always give similar results, as do the schemes LDA-SIC
and SIC-SIC. The two schemes in the first group evaluate the LDA energy
functional on self-consistent (LDA or SIC) densities, while the two 
schemes in the second group evaluate the SIC energy functional on
self-consistent (LDA or SIC) densities. The formation of these two groups,
occuring in the same manner for each of the three corrections,
shows us that what matters for the final result is the energy functional 
evaluated once on selfconsistent densities, not the potential employed during
the iterations to selfconsistency. This observation gives support to common 
post-LDA implementations of complex functionals.

For EFASIC and LESIC, the implementations in group two perform much worse than
those in group one. Within group one, SIC-LDA does slightly worse than
LDA-LDA, while within group two SIC-SIC is slightly worse than LDA-SIC. 
For all $U/t$, a simple selfconsistent LDA calculation is better than a 
calculation employing EFASIC or LESIC for the potential, the energy, or both.

For PZSIC, a crossover between the error curves occurs around 
$U\sim 5t$. Below this value, the picture is the same, qualitatively, as for 
LESIC and EFASIC: it is better to do a simple LDA calculation than to apply
PZSIC in any implementation. Above $U\sim 5t$, the situation changes, and the 
schemes in group two (LDA-PZSIC and PZSIC-PZSIC) perform clearly
better than those in group one.\cite{footnote3} Hence, only PZSIC is a real
improvement on the LDA, and this improvement manifests itself only for
sufficiently strongly interacting systems. 

As the data show, for weakly interacting systems (where the physics is 
dominated by the kinetic energy term, and the particles are delocalized)
LDA is still better than PZSIC. Of course, PZSIC removes the self-interaction 
error of the LDA also in this regime, but PZSIC itself is not
exact, and thus introduces other errors. For weak interactions, it does
not pay to correct a small error by a correction introducing larger
ones. For stronger interactions, the particles start to localize,
the SIE becomes more important, and its removal by PZSIC improves the total 
energies considerably. The separation in these two regimes is a very
gratifying result to obtain, as $U\sim 5t$ is known to mark the region where 
the Hubbard model crosses over from weakly to strongly interacting systems,
and the electrons become increasingly localized. The performance of PZSIC is 
clearly correlated with this crossover. In fact, in retrospect, one could have 
predicted that the electrons in the Hubbard model start to localize around 
$U \approx 5t$ simply by comparing the relative performance of LDA and 
LDA+PZSIC as a function of $U/t$.

To make this division in two distinct regimes clearer, we report in 
Table \ref{sictable} the rms error for each regime, {\em i.e.}, sum 
not only over systems with different size and particle number, but also 
over those with interaction strengths in the indicated range. The lowest rms 
error in each regime is printed in boldface. Clearly, for weakly interacting 
systems, LDA-LDA does best, but PZSIC-LDA and EFASIC-LDA are only marginally
inferior because the SIC chosen for the potential does not matter much,
compared to the effect of the energy functional.
For strongly interacting systems, on the other hand, LDA-PZSIC wins, closely 
followed by PZSIC-PZSIC. Overall, {\em i.e.} considering the rms error over
all considered systems, regardless of the value of $U/t$, LDA-PZSIC is the best
combination of correction and implementation among all methodologies tested 
here.

\begin{table}
\caption{\label{sictable} Root-mean-square error for all weakly and all 
strongly interacting systems (columns 1 and 2) and for all investigated systems
(column 3), for all ten permutations of methodologies investigated here.
Entries in boldface are the best in each column (interaction regime).}
\begin{tabular}{cccc}
\hline \hline
Method  & $U/t=1..5$ & $U/t=6..10$ & $U/t=1..10$  \\
\hline
$N_S$   & 210 & 210 & 420 \\
\hline
LDA-LDA &   {\bf 1.6008} &    4.2957 &    3.2416 \\
\hline
LDA-EFASIC  &  4.3713	 &  7.3541 &  6.0494  \\
EFASIC-EFASIC   & 4.4266 &  7.4423 &  6.1230 \\
EFASIC-LDA   & 1.6551 &  4.3827 & 3.3126 \\
\hline
LDA-PZSIC  & 2.6724 &   {\bf 3.0596} &  {\bf 2.8726} \\
PZSIC-PZSIC    & 2.7471 & 3.2693  &  3.0189\\
PZSIC-LDA   & 1.6567 &  4.4749 & 3.3741\\
\hline
LDA-LESIC   & 7.3476 &   13.9210 &   11.1306  \\
LESIC-LESIC   & 7.4749 &  14.3770 &  11.4580   \\
LESIC-LDA   & 1.7020 &  4.5960 & 3.4655 \\
\hline \hline
\end{tabular}
\end{table}

We finally remark that the peculiar aspect of the LE approach, where the 
SIC potential is neither the density derivative nor the orbital derivative
of the SIC energy, but a separate construction, does not seem to cause any 
additional features in the results. As we found above, the effect of the 
potential of any SIC is much less important than that of its energy, so the 
inconsistency between them in the LE approach apparently does not matter much. 

\section{Conclusions}
\label{conclusions}

We compared selfconsistent LDA calculations, three self-interaction 
corrections applied to the LDA, and three ways to implement each correction, 
for a total of 420 systems, for each of which we also obtained, numerically, 
the exact many-body solution. Three clear trends emerge from the resulting
4200 approximate and 420 exact data: 
(i) For all three investigated corrections, the main effect comes from the
energy, not the potential. Post-LDA implementations (correcting only the 
energy) and selfconsistent implementations (correcting the energy and the 
potential) give very similar results,\cite{footnote3} while implementing 
only the correction to the potential has virtually no effect on the resulting 
ground-state energies.
(ii) For weakly interacting systems, the benefit of an approximate SIC is 
overcompensated by the intrinsic errors of the approximation. Uncorrected 
LDA here works best.
(iii) For more strongly interacting systems, correcting the SIE is important,
but of the three investigated corrections only PZSIC systematically improves
on the LDA. Both the simple EFASIC and the more sophisticated LESIC are 
much worse.

Of course, all our conclusions may have been biased by the choice of the
Hubbard model as theoretical laboratory. The disadvantages of such possible
bias, on the other hand, are largely compensated by the availability of an
exact many-body solution as a benchmark, of the exact LDA,\cite{footnote2} and
by the absence of any problems due to finite basis sets or due to experimental 
uncertainties. It remains to be seen if the superiority of PZSIC remains 
unchallenged by extension of these investigations to other corrections, 
implementations and classes of systems. Similarly, the important issue of the
many-electron SIE \cite{msie1,msie2} requires separate investigation.

{\it Acknowledgments}
This work was supported by FAPESP and CNPq. KC thanks J. Terra and 
D. Guenzberger for bringing the LE approach to his attention. We thank
V.~L. Campo Jr. for his exact diagonalization code for the Hubbard model.

\end{document}